\documentstyle[12pt]{article}
\newcounter{num}

\title{A Classification Scheme for Toroidal Molecules}
\author{Jorge Berger and Joseph E.~Avron \\
Department of Physics, Technion -  Israel Institute
of Technology \\
 32000 Haifa, Israel\\
 e-mail: PHR76JB@VMSA.TECHNION.AC.IL }

\begin{document}

\maketitle

\section{Introduction}
The last decade witnessed the discovery of diverse hollow
graphitic structures \cite{Dov}. Several spherical
\cite{Fullerenes} and cylindrical \cite{Tubules} structures have been
observed. A natural question is
whether structures with other topologies, such as tori, also occur. It is
reasonable to expect that molecules that are multiply connected and have
negative curvature (i.e. have regions that approximate a saddle) would have
distinct physical and chemical properties.  For example,
certain quantum and transport phenomena depend on multiconnectivity
\cite{Rieman}, and it
is known that negative curvature has important  consequences for the
dynamics on the surface \cite{Chaos}.  Toroidal molecules may have
applications in host-guest chemistry \cite{hostguest}, where
the ``hole'' in the torus accomodates the ``guest". It
has also been suggested that toroidal molecules may have applications
as components in nanotechnology \cite{torus} and as catalizers
\cite{Ijima}.

Structures with two concentric cylinders
that inter-connect have been observed \cite{Ijima}.
Gluing two of these objects together would make a torus.
In the present study we shall focus on structures in the shape of a
doughnut.  Miscellaneous proposals for toroidal molecules have been
suggested \cite{torus,hexagons,Chern,Dunlap,azul}. Ab initio calculations
for the energetics and geometry of several toroidal carbon molecules have
been performed in \cite{first}.

The purpose of the present study is to present a classification
scheme for a wide class of toroidal molecules. This classification scheme
can distinguish between toroidal isomers. For example, Table I shows two
C$_{120}$ and two C$_{240}$ isomers of toroidal carbon that are resolved by
our classification.

Our tool is a set of tiling rules \cite{Tiling}, adapted to the
description of  noncrumpled toroidal molecules that approximate a surface
of revolution. These tiling rules reflect  the energy penalty for
stretched bonds and deformed bond angles. Our method is
amenable for computarized search (as the ``orange-peel" method \cite{Peel}
in the case of spheroids). A tiling pattern is determined by four integer
indices which characterize the basic ``unit cell", which is repeated
periodically to make a torus.
The approximate positions of all the atoms in the molecule
are determined by these four integers and by the number of unit cells in
the torus.
An announcement of some of the results in the present study is found in
\cite{A-B}.

Although we do not claim that all stable toroidal molecules
belong to the class of tilings that we introduce, this class contains
most toroidal molecules proposed in the literature and is large enough to
include close approximations to tori with prescribed reasonable
dimensions.

As an application we describe results of a numerical study of the energetics
of several toroidal carbon molecules. The study involved:
\begin{enumerate}
\item The relaxation of model carbon atoms  represented by tiling to
a stable molecular configuration, through  minimization of the
$\sigma-$bonds geometric potential energy. We have used  a
phenomenological potential due to Tersoff \cite{Tersof}. From here,
average energies and binding energies of the least bound
atoms were obtained.
\item HOMO-LUMO  and other properties of the $\pi$-electrons in
H\"uckel model.
\end{enumerate}

 The results (Table I)  support the conjecture that many
toroidal carbon  molecules are stable configurations.

\section{Tilings}
We are interested in surfaces that approximate toroidal surfaces of
revolution, as in
Fig.~1. (This surface is known as the ``anchor ring".) A point $P$ on this
surface has the cartesian representation
\begin{eqnarray}
R\left\{\left(1-\frac{\cos\theta}{\eta}\right)\cos\phi \ ,
\left(1-\frac{\cos\theta}{\eta}\right)\sin\phi \ ,
\frac{\sin\theta}{\eta}\right\}  \ ,               
\end{eqnarray}
where $\{\theta,\phi\}$ are the toroidal coordinates.  The parameter
$R>0$ determines the size, and the parameter $\eta > 1$ determines
the shape.  (Large $\eta$ corresponds to thin tori.)  Eq.~(1)
induces the metric and area \cite{geom}
\begin{eqnarray}
&&\eta^2 R^{-2}ds^2=d\theta^2+(d\phi)^2  \
(\eta-\cos\theta)^2       \ ; \nonumber \\       
&&\eta^2 R^{-2}dA = d\theta d\phi (\eta-\cos\theta) \ .
\end{eqnarray}
A toroidal surface may be represented by the rectangle $[0,2\pi] \times
[0,2\pi]$ in the ($\theta$,$\phi$) plane.
Opposite sides of this rectangle are identified (i.~e., are understood
as being the same line); this identification is equivalent to the
periodicity of Eq.~(1) in $\theta$ and $\phi$. We shall refer to lines of
fixed $\theta$ (resp. $\phi$) as
latitudes (resp. longitudes) and draw them horizontal (resp. vertical).

 A toroidal carbon molecule can be represented by a graph on such a rectangle
(like the solid lines in Fig.~2), where  atoms sit at vertices and each
vertex is trivalent.
(Following the nomenclature of graphs theory, the {\it valence of a vertex}
is the number of edges that meet at the vertex; if atoms sit at the
vertices this is the same as the chemical valence.)
 Since carbon bonds prefer angles near $120^\circ$, graphs
with pentagonal, hexagonal and heptagonal rings are favored and we allow
only these.

It is known that pentagonal rings are associated with positive (Gaussian)
curvature and heptagonal rings with negative curvature. Tori have positive
curvature far from their axis (where their surface is locally similar to
an ellipsoid) and negative curvature in the hub region (where the surface
looks like a saddle). Therefore, noncrumpled toroidal molecules should
contain pentagonal and heptagonal rings (in equal numbers by Euler's
Theorem). The integrated Gaussian curvature of the torus in Fig.~1 for
$-\pi/2 < \theta < \pi/2$ is $-4\pi$, and is cancelled by the integrated
Gaussian curvature over the rest of the torus. Since the positive part is
$4\pi$, just like for the sphere, it suggests that the number of
pentagons should be about twelve (and equal to the number of heptagons).
Indeed, our numerical study shows that hexagons-only
tori (e.~g. \cite{hexagons}) have large energies; they have atoms that
are strained about three or four times more than the most strained atoms
in typical molecules that allow for pentagons and heptagons.

It will be useful to represent a molecule by its {\it dual graph}.
The dual graph is constructed by drawing a vertex inside every face of
the direct graph, and connecting vertices of faces with a common edge.
The dual graph of Fig.~2 has been drawn with dashed lines. The duality
operation exchanges faces (and their number of edges) with vertices (and
their valence). Since we are considering trivalent atoms, and since we
allow rings with 5, 6 and 7 atoms, the dual graph
consists of triangles and has vertices of valences $5,\ 6$ and $7$.
(The name ``dual" reflects the fact that the dual of the dual
is isomorphic the initial graph.)

 We shall idealize the triangles in the
dual graph and represent them by {\it tiles}: by this we mean that, instead
of having all sorts of shapes, only a small family of triangles will be
used as building blocks of our dual graphs. Our tiles will be
right-angled isosceles triangles, as in Fig.~3; the motivation for this
choice is given in the appendix. As discussed in the appendix, our
 set of tiles includes various sizes,
reflecting the non uniform metric in the $\{\theta,\phi\}$
plane.  (Large tiles are placed where $\cos\theta$ is large and vice
versa.) It follows by Pythagoras that the different tiles are related by
scaling by powers of $\sqrt{2}$.
Due to the privileged role of latitudes and longitudes, we shall (initially)
require that each tile have at least one side along one of these directions.
More general structures  will be discussed in Section 5.

Tiles will be classified by their sizes, and assigned a {\it generation
number}.  The largest tiles will be called the
 first generation and the tiles in the  $i$-th generation
are those with an area which is $2^{i-1}$ times smaller.
In the first generation we
allow only the two orientations (related by a $\pi$ rotation)
with the hypotenuse along a latitudinal line as shown in Fig.~3a;
 this will be required from every odd generation.
This turns out to give rise to an array of equidistand horizontal lines, a
feature that agrees with metric (2), which
depends on $\theta$, but not on $\phi$.

In the second generation we allow the four
 tiles in Fig.~3b.  Tiles in generation $p+2$ are the same as
those in generation $p$, except for scaling by a factor
of $\frac{1}{2}$. Clearly, tiles that share a common edge
differ by at most one generation.
Since there is only one way of matching
odd generation tiles, the tiling of an odd-generation region
is uniquely determined by its border; on the other hand,
since a square can be tiled in two ways (see Fig.~4), the tiling
of an even-generation region
involves a choice of orientation for each square.

We define the {\em length} $L(\theta)$ of a latitudinal
line as the number of edges it contains. Our tiling rules have the
property that
$L(\theta)$ is larger when the line passes through high generation regions.
In Fig.~5, $L(\theta)$ varies from 1 in the horizontal lines in the
first generation to 3 in the third-fourth generation region.
Likewise, we define the length of a longitudinal line as the number
of horizontal lines it intersects.  Our tiling rules ensure that the
total length of longitudinal lines is independent of $\phi$, as it
should.  This length will be dubbed the ``girth'' and denoted by
$g$.  We also define $g_i(\phi)$ as the length within the $i$-th
generation regions.  If a longitudinal line crosses
several regions of the same generation $i$, then $g_i(\phi)$
is defined as the {\em total} length across all these regions.
In Fig.~5, $g_1 =g_3 =2$ and $g_2 =g_4 =1$ (independent of $\phi$).
Denoting by $m$ the number of generations in the tiling,
 $\/ \sum^m_{i=1} g_i(\phi)=g$.\\

Since the linear dimension of the last generation is smaller than
that of the first by a factor $2^{(m-1)/2}$, and since the first generation
has the hypotenuse along a latitudinal line (while the last generation
has this property only if $m$ is odd, introducing
an additional factor $2^{1/2}$ in the counting metric along the latitudinal
direction when $m$ is even), Eq.~(2) can be used to relate $\eta$ to the
number of generations $m$:
\begin{eqnarray}
\frac{\eta+1}{\eta-1} \sim 2^{[m/2]} \ ,    
\end{eqnarray}
where $[m/2]=$ integral part of $m/2$.
Therefore, the
number of generations plays a role analogous to $1/\eta$.
A cylinder can be built with one generation,  skinny tori
 are associated with first and second generation tiles and, the fatter
the torus, the more generations are needed.

Unlike the sphere, which is unique up to scaling, the tori of Eq.~(1)
depend also on the shape parameter $\eta$. {\it A priori}, it is not clear
whether this family has a distinguished member. It turns out that
the case $\eta = \sqrt{2}$ is special. In the theory
of surfaces the torus of revolution with $\eta = \sqrt{2}$ is known as the
Clifford torus and is distinguished by being the minimizer of the
Willmore functional, i.~e., it minimizes the square of the mean curvature
\cite{Bensimon}.
The fattest torus we shall study has five generations, and is inspired by
the Clifford torus. As we shall see, it does have marked stability.
For fatter tori, $\eta < \sqrt{2}$,
axially symmetric structures are disfavored \cite{Bensimon}.

As mentioned above, valence plays the role of
curvature: valence 5 (resp. 6, 7) vertices carry positive (resp. zero,
negative) curvature. Since curvature also implies non-constant metric,
we want no 5 and 7 valence vertices embedded in single generation
regions, where all triangles have the same Euclidean area. The pair of
tiles of the odd generations have
indeed this behavior built into them:  only sixfold vertices can be made
with them alone.
On the other hand, Fig.~4 shows a tiling by  even
generation tiles with valences 5 and 7.
These pentagon-heptagon pairs do not produce curvature of the form
required by metric (2); they merely produce crumpling, and their
integrated effect within a single generation region is zero.
Therefore, pentagons or heptagons will be allowed
only at the interfaces between different generations.

  The predominance of pentagons over heptagons is the
discrete analog of the (Gaussian) curvature.  In our case, it should be
a decreasing function of $\cos\theta$ and, also, of the size of the
tile. Therefore, the predominance of pentagons over heptagons ought to be
an increasing function of the generation number.
However, since every pentagon-heptagon pair necessarily involves angles
that differ from 120$^\circ$, we shall attempt to include as few of them
as reasonable. Moreover, we shall deal only with small numbers of
generations ($\le 5$), so that it seems reasonable to
compromise the smoothness of the curvature on behalf of limited
numbers of pentagons and heptagons, and locate them only where
they are most required. In summary, we impose the following:\\

\noindent {\em 5-6-7 Rule}: For a torus with $m$ generations, all the
heptagons lie at the interfaces between the 1st and the 2nd generation
and all the pentagons lie at the interfaces between the $m$-th and the
$(m-1$)-th generation.  (If there are only two generations,
both the heptagons and the pentagons lie at the interfaces between
them.)  Furthermore, we require that around every heptagon (resp.
pentagon) there be a majority of tiles of generation 1 (resp. $m$).\\

Figs. 5-7 are examples of tilings which obey the 5-6-7 rule.

\section{Skeletons and Relationships}
Given a set of interfaces between generations, the position of each tile
in the tiling is determined uniquely by the 5-6-7 rule (although not for
every set of
interfaces does a legal tiling exist).  The set of interfaces will be called
the {\it skeleton}.  The ($2 p-1,\ 2p)$ interfaces are zigzag lines (e.g.
Fig.~8a) and the $(2p, \ 2p+1$) interfaces are horizontal lines
(e.g. Fig.~8b).  Therefore, regions that contain $2p-1$ and $2p$
tiles lie within a horizontal stripe, which will be called a
{\em $p$-th stripe}.  (If there are only 2 generations, there is just
one stripe; if the number $m$ of generations is odd, then the
last stripe has tiles of the $m$-th generation only.)
Figs.~6 and 7 have one stripe; Fig.~5 has two.

Since according to metric (2) the length of the latitudinal
lines has only one minimum and one maximum, we require that the
regions of generation $i$, $1\leq i\leq m$, be located in monotonic
and consecutive order, both when going from region 1 to $m$ by
increasing or by decreasing $\theta$.  The 5-6-7 rule imposes a
strong constraint on the class of possible skeletons with two or more
stripes: all the zigzag interfaces are parallel to each other and,
 with the exception of the pentagonal and possibly the heptagonal
vertices,  the vertices of the zigzag interfaces lie at the stripe
borders (see Fig.~9).  This has
two consequences: First, the tiling has mirrors of
symmetry at the longitudinal (vertical) lines that contain the
pentagons and the heptagons; it also has centers of inversion that lie
midway
between the heptagons (resp. pentagons) which are located at different
zigzag lines (keeping periodicity in mind).  Second, the $g_i$'s are
independent of $\phi$, in agreement with the our target metric (2).
  If the skeleton consists of just one stripe, then the
symmetry, the constancy of the $g_i$'s and the zigging
between two lines of constant latitude don't follow from the 5-6-7
rule (e. g. Figs.~10-11), but we impose them in order to mimic the
symmetry properties of metric~(2).
The independence of the $g_i$'s on $\phi$ rules out the tori
considered in Refs. \cite{Chern,Dunlap}, which are cylindrical tubes
connected by elbows.

We are thus left with a class of tilings such that the skeleton
(and the entire tiling) is completely determined by four numbers (see
Fig.~9):
$m$ (the number of generations), $g_1$ and $g_m$  (the contributions
of the first and last generation to the girth), and $z$ (the number of
edges in a zig of the (1,2) interface). For a generation number $i$ which
is odd, $g_i$ must be even.  If $i \ne 1,m$, then
$g_i=2^{p-1}z$, where $p$ is the number of the stripe.

The girth of the torus is given by
\setcounter{num}{4}
\setcounter{equation}{0}
\def\theequation{\thenum\alph{equation}}
\begin{eqnarray}
g=g_1+g_{\rm mid} + g_{\rm end} + g_m \ ,                
\end{eqnarray}
with
\begin{eqnarray}
g_{\rm mid} = \left\{
        \begin{array}{ll}
        0 & m \leq 2 \\
        z & 2<m<5 \\
        (2^k-3)z & m \ge 5
        \end{array} \right. 
\end{eqnarray}
where $k=[(m+1)/2]$ is the number of stripes ([ ] denotes integral part),
and
 \begin{eqnarray}
 g_{\rm end} = \left\{
        \begin{array}{ll}
        2^{k-1}z &  m \mbox{ even and} \ge 4 \\
        0 & \mbox{otherwise}
        \end{array} \right. 
\end{eqnarray}

The $\phi$-range spanned by a zig and a zag will be called a {\em unit cell}.
The number of atoms in a unit cell is given by
\setcounter{equation}{4}
\def\theequation{\arabic{equation}}
\begin{eqnarray}
A = \left\{
       \begin{array}{ll}
       2z\left(g_1+2^{[m/2]}g_m\right) & m<3 \\
 2z\left(g_1+2^{[m/2]}g_m+(2^{m-1}-2)z\right)  & m \ge 3
        \end{array} \right. 
\end{eqnarray}

The length of the shortest latitudinal line in a unit cell is given by
\begin{eqnarray}
L_{\rm min}=2z-{\rm Min}(g_1,z) \ . 
\end{eqnarray}
The length of the longest latitudinal line in a unit cell is
\begin{eqnarray}
L_{\rm max}= \left\{
       \begin{array}{ll}
       2^{k-1}z & m \mbox{ odd} \\
       2^{k-1}z+{\rm Min}(g_m, 2^{k-1}z) & m \mbox{ even}
       \end{array} \right.   
\end{eqnarray}

A unit cell contains 2 pentagons and 2 heptagons.

\section{From Tiles to Tori}
We build a torus in three dimensional space from a latitudinal array of
$n$ unit cells. By associating periodic coordinates $\{\theta,\phi\}$ to the
tiling plane, the three-dimensional positions follow by Eq.~(1).
The line $\theta=0$  is assigned to the horizontal
line through the centers of inversion near the vertices of valence
7 (such as the ``X" in Fig.~5) and the horizontal line $\theta=\pi$
passes through the centers of inversion near the vertices of valence 5
(such as the asterisk).  We then increase (resp. decrease) $\theta$ by
$2\pi/g$
when going up (resp. down) by one horizontal line.  $\phi$ is assigned
by requiring it to span $2\pi$ in the periodic tiling and to be linear
along the horizontal planar distance in the tiling plane.  Finally,
we impose periodicity $\{\theta+2\pi,\phi\} \equiv \{\theta,\phi+2\pi\}
\equiv \{\theta,\phi\}$. For example, for the molecule obtained
by 5 repetitions of the unit cell shown in Fig.~5, if the longitude
$\phi=0$ is chosen to
pass through the ``X" and the asterisk, the coordinates of the center of the
disk (located at $\frac{1}{3}$ of the height of the triangle) would be $\{
\theta=19\pi/18,\phi=-\pi/10\}$.

{}From the symmetry of the tiling, discussed in the previous section, it
follows that a torus with $n$ unit cells has symmetry D$_{nd}$, which is
the highest symmetry of a torus in three dimensions that could be
expected from a discrete construction.

 The number of unit cells
 in the molecule may be estimated a priori by comparing the difference in the
perimeters at $\theta = 0$ and $\theta=\pi$ to that of a geometric
torus.  This gives
\setcounter{equation}{7}
\def\theequation{\arabic{equation}}
\begin{eqnarray}
n \sim n_0 = 2 g/(L_{\rm max}-L_{\rm min}) \ ,                    
\end{eqnarray}
where $g$, $L_{\rm max}$ and $L_{\rm min}$ are given by (4), (6) and (7).

To evaluate the parameters $R$ and $\eta$ which
appear in Eq.~(1), $R$ (resp. $R/\eta$) is taken proportional to the
average latitudinal (resp. longitudinal) length, and the constant of
proportionality is a ``material parameter". For carbon atoms,
appropriate values are
\begin{eqnarray}
 R=0.2n(L_{\rm min}+L_{\rm max}) {\rm \AA} 
\end{eqnarray}
and
\begin{eqnarray}
R/\eta=0.4g {\rm \AA} \ .          
\end{eqnarray}
(The geometric data in Table~I do not follow directly from Eqs.~(9) and
(10); they were obtained
after refinement through minimization of the interatomic potential.)

The present section suffices to deal with most molecules considered in
this paper. Section~5 provides generalizations, and may be skipped
on first reading.

\section{Associated Tilings}
Each of the tilings of the previous sections can be used to generate
several descendent structures. These are more general than those considered
so far and in some cases involve tiles of different shapes.
The generalizations we consider are Goldberg
inclusion, elongation, chirality and torsion.

{\em Inclusion} consists of drawing $v$ vertices inside each
tile and then connecting both the old
and the new vertices to form a new array of triangles in the dual graph.
Vertices on the edge between two tiles are counted as $\frac{1}{2}$
in each. This procedure increases the number of atoms by a factor
$q=1+2v$. Connections are made so that old vertices
retain their valences, new vertices have valence 6, and nearest vertices
are connected. We restrict ourselves to the cases in which this rule for
the connection of the vertices is unambiguous.
If $q=1+2v$ is of the form
\setcounter{num}{11}
\setcounter{equation}{0}
\def\theequation{\thenum\alph{equation}}
\begin{eqnarray}
q=\ell_1^2 +\ell_2^2 +\ell_1 \ell_2  \ , 
\end{eqnarray}
with $\ell_1$ and $\ell_2$ integers, then a distribution of vertices for
which connections are unambiguous is obtained by a variation of
the {\em Goldberg transformation} \cite{Gold}: using local
coordinates for each tile, such that its vertices are at $(x=0, \ y=0)$,
$(x=1, \ y=0)$ and $(x=0, \ y=1)$ (with {\bf \^{x}} to the right of
{\bf \^{y}}), the included vertices are at
\begin{eqnarray}
x &=& (n_1 (\ell_1 +\ell_2 )+ n_2 \ell_2)/q \ , \nonumber
\\                    y &=& (-n_1 \ell_2 +n_2 \ell_1 )/q  \ ,   
\end{eqnarray}
where $(n_1 ,n_2 )$ are integers such that $(x,y)$ lies inside the tile.
Fig.~12 shows the Goldberg transformation $(\ell_1 ,\ell_2 )=(1,2)$
applied to the tiling $g_1=2, \ g_2=z=1$.

 Inclusion generates a family of
molecules for which the number of pentagons and heptagons and their
relative positions (and therefore the overall shape) remain fixed.
Except for the cases $\ell_1 = \ell_2$ and
$\ell_1 \mbox{ or } \ell_2 = 0$, which are considered below, the Goldberg
inclusion gives rise to chiral molecules.

A Goldberg inclusion with indices $(\ell_1 ,\ell_2 )=(\ell,0)$ (or
$(0,\ell)$) will be called {\em inflation}.
 Inflation is the discrete analog of scaling. Inflation
by  a factor $\ell$ replaces  all the tiles by  ``super-tiles"
where each new tile is made of $\ell^2$ scaled copies of the old tile
(or its
$\pi$ rotated image).
Inflation  preserves the generation number, increasing by the
factor $\ell^2$ the number of atoms in each.
Inflation by a factor $\ell$ gives a tiling within the class we have
considered, leaving $m$ and $n$ unchanged, while leading to
multiplication of $z$ and all the $g_i$'s by $\ell$.  Therefore,
for classification purposes it is convenient to consider primitive
tiling, i.e. sets of numbers $(g_1,g_m,z)$ which have no common
divisor.  (Except possibly 2, since $g_i$ must be even for odd $i$.)

A Goldberg inclusion with indices $\ell_1 =\ell_2 =1$ is
a {\em leapfrog} \cite{Leapfrog}.
In three dimensional space, the leapfrog
operation is a vertex truncation through the mid-points of the edges, so
that every old vertex turns into a hexagon and the number of atoms is
multiplied by 3.  Leapfrogged structures are guaranteed to
have closed shells in the H\"{u}ckel spectrum \cite{MWF}.
The effect of leapfrogging twice is the same as that of inflation by a
factor 3.

The tori of Ref. \cite{Chern} may be obtained by {\em elongation}.
Elongation is defined for tilings with two generations. It consists of
cutting the tiling along an approximately longitudinal line (following
the edges of the tiles), separating the two pieces by an integer
latitudinal distance, and filling in with tiles of generation 2, in such
a way that all additional vertices are hexagonal. Fig.~10 shows an
elongation of the tiling $g_1=2, \ g_2=z=1$.

The tori considered by Dunlap \cite{Dunlap} have tilings which typically
look like that in Fig.~13, and do not obey the 5-6-7 rule. These tori
have D$_{nh}$ rather than  D$_{nd}$ symmetry. Their curvature does
not vary smoothly as in metric~(2); it concentrates at the defects.

For many tilings there exist latitudinal lines  $\theta=\theta^*$ which
are embedded in a single odd generation (first or last, with
$L(\theta^*) \equiv L^*$ equal to either $nL_{\rm min}$ or $nL_{\rm
max}$). If this
is the case, manifest {\em chirality} \cite{Dress} may be introduced by using
$\theta$ in the range $\theta^* \leq \theta < \theta^*+2\pi$ and
associating to the coordinates $\{\theta,\phi\}$ in the tiling the
coordinates $\{\theta,\phi+j\theta/L^*\}$ in three dimensional space,
with $j$ integer. (This is the analog of Dehn twist in the theory of
Riemann surfaces.)
 This amounts to identification of the edge at
$\theta=\theta^*$ with the edge at $\theta=\theta^*+2\pi$ shifted by $j$.
The 5-6-7 rule will still be fulfilled. Unless $|j|<<L^* $ and $|j|<<g $,
this shift produces considerable strain. It is possible to construct tori
with $g_1$ odd or with $m$ odd and $g_m$ odd, provided that we introduce
a chiral shift by a half-integer $j$.

Shifts in the longitudinal boundaries of the tiling  would lead in general
to a misfit in the skeleton. However, if there
are only two generations, {\em torsion} can be introduced by drawing zigs
($z_1$) longer than the zags ($z_2$), as in Fig.~11. The resulting
molecule is an open helix rather than a torus. The sinus of its pitch
angle may be estimated as $2(z_1 -z_2 )/(L_{\rm max}-L_{\rm min})$. Helical
structures have been studied in \cite{helices}.

Let us mention that better approximations to metric~(2)
can be achieved by two-generation tilings in which the skeleton consists
of two parallel jagged lines, such that the number of heptagons minus the
number of pentagons in the range $[0,\theta ]$ is approximately
proportional to
$|\sin \theta|$. Tilings like this would have a very large number of atoms
per unit cell and, also, a very large $n_0$ in Eq.~(8).

\section{Energetics}
For each molecule, we calculate two complementary
phenomenological contributions to the energy. The geometric contribution
is calculated by means of an interatomic potential, and the
contribution of delocalized electrons by means of H\"{u}ckel model.

\noindent {\it Interatomic Potential}: We use Tersoff's potential
\cite{Tersof}.

The tiling rules of the previous
sections not only describe the connectivity of the molecule, but also
suggest values for  $n$, $R$ and $\eta$ through Eqs.~(8)-(10). The carbon
atoms (located at the centers of each triangle), are assigned
initial positions in space which correspond to their toroidal
coordinates.  The positions of all the atoms are then allowed to vary
in three-dimensional space to minimize the total potential energy of
the bonds. Good tilings flow to a nearby local minimum. Uncontrolled
guesses for the initial coordinates result in atoms being lost to infinity.

We have used Powell's method \cite{Powel} to flow
to a local minimum of the molecular
energy. In all cases, our tilings provided initial
guesses for the atomic positions within the basin of convergence of the
minimization algorithm. When the number of coordinates is large, analytic
expressions for the derivatives of the energy must be supplied in order
to obtain reliable minima.

For every tiling, we tried several values for the number $n$ of unit cells.
For the tilings considered, the largest binding energies of the entire
molecule were
usually obtained for $n=6$ and the largest binding energy of the most
strained atoms, for $n=5$.
The results  are shown in Table I.  Since the differences
between $n=5$ and $n=6$ are small, we present the results for $n=5$ only.
To facilitate
comparison, Tersoff's energies are given relative to that of an ideal
graphitic plane, which is $-7.40$ eV per atom.  It is instructive to
compare these energies with those of a torus with girth 4 and 160
atoms, but with hexagonal rings only: in this
case $N(\bar{E} - E_G)/|E_G|=29$ and $10^3(E_{\rm worst} - E_G)/|E_G|=411$
(and strain increases for larger girth).

\noindent {\em H\"{u}ckel Model}:
H\"uckel's model \cite{Graph} associates an $N\times N$  incidence matrix
to a
graph of $N$ carbon atoms, whose  eigenvalues, $E_j,\ j\in\{ 1,\dots,N\} $
correspond to the discrete energies of independent delocalized
$\pi$ electrons.
A closed shell, where  half the eigenvalues
are negative and half positive ($N$ is even) implies
stability  (protecting e.g. against Jahn-Teller instability).
The largest occupied eigenvalue, $E_{N/2}$, is known as {\it HOMO}-energy
and the smallest unoccupied one, $E_{N/2+1}$, as {\it LUMO}. A large gap
is a criterion for stability. Taking $\frac{1}{N} \sum E_j$ as the reference,
$|E_j|\le 3|\beta|$, where $\beta$ is the nearest-neighbor overlap integral.

A computer program which calculates the H\"{u}ckel
spectrum for arbitrary $g_1,g_2,z$ and $n$ is available upon request.
 Illustrative
values  are summarized in Tables I and II.  The average energy
per atom is almost insensitive to the number $n$ of unit cells.

The H\"uckel model also provides an
estimate for the electronic charge distribution in the torus.
 The general trend is that five (resp. seven)-atoms rings
capture (resp. release) electrons from/to their adjacent atoms. The total
capture/release is of the order of a quarter of an electron per ring.

\noindent {\em Choice of Examples}: The method described above allows
to design and analyze infinitely many toroidal molecules.
One can for instance require the inner radius of the torus
(which is proportional to $nL_{\rm min}$) to fit the size of a
given ``guest"; given the mass of a molecule,
one could fix $N/n=A$ in Eq.~(5); in order to design an elbow
that fits a given tubule or cap, one would require a given girth.

The case of two connected concentric cylinders \cite{Ijima} may be
reproduced by a two-generation torus with $g_1 \approx g_2 >> z$. Since
experimentally found ``turn around" edges have a small difference between
the external and internal radius (relative to the radii themselves),
elongation as in Fig.~10 is necessary. The outer cylinder contains tiles
of generation 2, connections contain the zigzagging boundaries, and
the inner cylinder contains tiles of generation 1 and separating
fringes of generation 2. From geometric arguments and
comparison with Eqs.~(9) and (10), we estimate $g_1$ and $g_2$ as $ 0.4
\times ($length of the cylinders), $nz$ as $2.5 \times ($difference in
radii of cylinders) and the total horizontal witdth of the separating
fringes as $5 \times ($the internal radius) minus $2.5 \times ($the
external radius), with all lengths in \AA.  Note that in this case
Eq.~(1) is not a good approximation
and that interlayer interaction has not been taken into account.

The primitive tori ${\rm C}_{120}$ and ${\rm C}_{240}$ in \cite{torus}
may be traced to be $g_1 =2, \ g_4 =z=1$ and the leapfrogged of
$g_1 =g_3 =2, \ z=1$, respectively. The smallest possible tiling, $g_1
=2, \ g_2 =z=1$ is an azulenoid \cite{azul}.

For the purpose of illustration, we have focused on small values for
$g_1,g_m$ and $z$. Molecules with longitudinal
perimeters of the order of that of C$_{60}$ imply girth $g \sim 10$.
For tori with roughly circular cross section, we may also expect that the
difference
in the latitudes of the pentagon and the heptagon lying at the same
longitude be at least about $2\pi/3$.

Fig.~14 shows three dimensional views of the two-generations molecule
$m=2, \ g_1=g_2=z=4$
after relaxation to the minimal energy, and Fig.~15 shows $m=5,\ g_1 =
g_5=2, \  z = 1$. The latter is the five-generations molecule with the
smallest
possible indices $g_1$, $g_5$ and $z$. As expected, the five-generations
torus is much fatter, and is close to a Clifford torus \cite{Bensimon}.

Comparison of different levels of inflation of $m=g_1=g_2=z=2$ shows that
the configurational energy per atom relative to graphite decreases
roughly as $1/N$. (Inflation is the discrete analog of scaling and was
discussed in Section~5; inflation by the factor $\ell$ consists of
multiplying $g_1$, $g_m$ and $z$ by $\ell$, and its approximate effect
is increase of linear dimensions by this factor.)
The same result was obtained in \cite{torus}.
  However, the energy of the least bound atoms saturates.
This effect is probably due to the fact that the number of pentagons and
heptagons is kept unchanged, and they bear the entire burden of
curvature.  In Fig.~14a it is clearly seen that pentagons stick out.
It is also interesting to note in Fig.~14b how the
pentagons deviate from the latitude $\theta = 0$ which they have in the
tiling.

Fig.~16 shows a chiral molecule, obtained by shifting by $j=1$ the tiling
of the molecule in Fig.~15 at $\theta^* = \pi $.

The influence of inflation on the H\"{u}ckel spectrum was studied for
$g_1=g_2=2$ and $1 \leq z \leq 3$. The results are summarized in Figs.~17
and 18. Fig.~17 describes the HOMO-LUMO gap. Since there are $N$ energy
levels, and since they are bound in the interval $[3\beta,-3\beta]$, the gap
might naively be expected to be inversely proportional to $N$ (and thus to
$\ell^2$, where $\ell$ is the inflation factor). The trend is,
however, that the gap is inversely proportional to
$\ell$. This result resembles the case of tubules, where the gap is
inversely proportional to the diameter \cite{diameter}. There are some
inflation values for which the regular trend is not obeyed. These are a
reminiscent of gap closure for tubules whose width is divisible by 3.

Fig. 18 shows how the delocalisation energy per atom approaches that of
graphite. The difference from graphite is, approximately, inversely
proportional to $\ell^2$.

\section{Conclusions}
Tori are closed surfaces, like spheres, but unlike spheres  are not
simply connected. They may provide an arena for a variety of physical and
chemical properties.

Table I shows that, from  the energetic point
of view, there are toroidal molecules with promising stabilities and
one could say that the fact that toroidal
carbon molecules have  not been observed is as significant as the fact
that C$_{60}$ molecules were not observed until recently.
It is difficult to draw definite conclusions from the
comparison of binding energies of molecules that are not isomers.
In general,
the binding energy per atom increases to the graphite limit as the
number of atoms increases and we have found, as in
\cite{torus}, that the total binding energy relative to
graphite, $N(\bar{E}-E_G)$, remains roughly constant when the
shape of the molecule is kept fixed and the size is increased by means of
Goldberg inclusions. $N(\bar{E}-E_G)$ is
therefore a representative measure of the price in energy required to
form a torus, and is listed in Table I.

 The configurational energies and the
geometries we have found are similar to those in \cite{torus}, in spite
of having used a different potential and minimization technique. Also,
the trend of 5-atom (resp. 7) rings to capture (resp. release) electrons
is of the same magnitude as that found in \cite{first}. The
H\"uckel model predicts that the molecule $g_1 =2, \ g_4 =z=1$ (known as
C$_{120}$) has degenerate HOMO-LUMO states; this result disagrees with
\cite{first}.

We have developed a classification scheme which can be used to ``design''
toroidal molecules with prescribed dimensions. As long as these molecules
are not found, it may help to suggest candidates; if they are found, it
may help to sort and analyze them.

{\bf Acknowledgments:}
We are grateful to D.~Zohar and M.~Kaftory for help with Figs.~1
and 14-16,
to M.~Dresselhaus, L.A.~Chernozatonskii, T.~Schlick and B.~Dunlap for
correspondence and to D.~Agmon for reading the manuscript.  We have
benefited from remarks and information provided by the referees. The
research is supported by  GIF and DFG through  SFB288, and the Fund for the
Promotion of Research at the Technion.

\noindent {\Large\bf Appendix: Choice of Tiles}

Not every set of triangles is acceptable for tiling. The
tiles must permit a covering of a rectangle with periodic boundary
conditions such that adjacent tiles have matching edges. Tiles must also
be compatible with vertices with valences 5, 6 and 7. Besides these
restrictions,
choosing tiles involves some arbitrariness. Our choice was guided by two
criteria: geometrical considerations and simplicity. Geometrical
considerations should in the end enable us to obtain molecules in which all
bonds have approximately the same length and form angles close to
120$^\circ$; simplicity will facilitate bookkeeping.

The metric in the tiling plane, Eq.~(2), is not Euclidean in
$\{\theta,\phi\}$. As a consequence, equal
areas in the tiling plane do not correspond to equal areas in the toroidal
surface. A unit area in the Euclidean metric of $\{\theta,\phi\}$ for
$\theta \sim \pi$ corresponds to a larger
area (and, therefore, a larger number of atoms) in the molecule than a
unit area for $\theta \sim 0$. Since every tile represents one atom,
more tiles cover the latitude for $\theta \sim \pi$ than for $\theta \sim 0$,
so that the former tiles must be smaller. We are thus forced by geometry
to mind tiles of different sizes; we use simplicity to decree that all tiles
be similar. This leaves us with one triangular shape to choose.

Since equilateral triangles are
incompatible with valences other than 6 and imply a single
tile size, the simplest possibility is that of isosceles triangles. The
angles $(\alpha,\beta,\beta)$ of the tile must fulfill
\begin{eqnarray}
    \alpha+2\beta=\pi \mbox{~~~~~~~~~~~~(A1)}  \nonumber
\end{eqnarray}
and the valence condition amounts to the three equations
\begin{eqnarray}
    n_j\alpha+(j-n_j)\beta=2\pi, \mbox{  }  j=5,6,7,
    \mbox{~~~~~~~~(A2)}  \nonumber
\end{eqnarray}
where $j$ stands for the valence of the vertex and $n_j$ is an integer
between 0 and $j$. Eqs.~(A1)-(A2)
have exactly three solutions: $(\alpha=\pi/5,n_5=0,
n_6=2,n_7=4)$, $(\alpha=\pi/2,n_5=3,n_6=2,n_7=1)$ and
$(\alpha=3\pi/7,n_5=4,n_6=2,n_7=0)$. We now use the requirement that
adjacent tiles have matching edges. Since both sides of $\alpha$ are equal
and both sides of $\beta$ are different, going around a vertex will take us
to the same edge only if $\beta$ is present an even number of times, i.~e.,
$j-n_j$ must be even for all values of $j$. The only solution that satisfies
this requirement is $\alpha=\pi/2$, and we are left with right-angled
isosceles triangles as in Fig.~3.

\newpage

\begin{center}
Table I
\end{center}

\begin{tabular}{lcccccccc}
$m$ & 2 & 2 & 2 & 2 & 3 &  4 & 5 & $5^1$ \\
$g_1$ & 2 & 4 & 6 & 2 &4 & 2 & 2 & 2\\
$g_m$ & 2 & 4 & 6 & 2 &2 & 1 & 2 & 2\\
$z$ & 2 & 4 & 6 & 3 &6 & 1 & 1 & 1\\
$g$ & 4 & 8 & 12 & 4 & 12 & 6 & 9 & 9 \\
$N$ & 120 & 480 & 1080 & 180 &1200 & 120 & 240
& 240 \\
$\epsilon_h(\beta)$ & 0.43 & 0.16 & 0.08 &0.45 &0.07 & 0.04
& 0.19 & 0.20\\
$\epsilon_\ell(\beta)$ & -0.27~ & -0.22~ & -0.02~ &-0.18~
&-0.02~ & 0.04 & 0.01 & -0.04~ \\
$10^3(\bar{\epsilon}-\bar{\epsilon}_G) (|\beta |)$ & 16 & 4 & 2 &11
&2 & 25 &11 & 9\\
$10^3(\bar{E} - E_G)/|E_G|$ & 123 & 32 & 16 & 118 & 16 &  76
& 39 & 43\\
$10^3(E_{\rm worst} - E_G)/|E_G|$ & 181 & 114 & 108 & 170 & 115 &
122 & 101 & 100\\
$N(\bar{E} - E_G)/|E_G|$ & 14.8 & 15.6 & 17.5 & 21.2 & 19.5 & 9.1
& 9.4 & 10.4\\
inner radius ($\rm \AA$) & 4.05 & 8.33 & 12.57~ & 6.87 &
14.15~ & 2.02 & 2.09 & 2.12\\
outer radius ($\rm \AA$) & 7.29 & 14.21~ & 21.17~ & 10.14~
&23.42~ & 6.02 & 8.49 & 8.61\\
height ($\rm \AA)$  & 2.93 & 5.68 & 8.00 & 3.37 & 8.13 & 4.57 & 6.89 & 6.96
\end{tabular}
\pagebreak

%

\begin{center} Table II \end{center}

\begin{tabular}{|cc|ccc|ccc|}  \hline
\verb+\+ & $g_2$ & 1 & 2 & 3 & 1 & 2 & 3 \\
$z$ & \verb+\+ & &$g_1=2$ & & & $g_1=4$ & \\      \hline
1 & & 35 & 17 & 22 & 36 & ~0 & ~1 \\
2 & & 61 & 70 & ~8 & 11 & 51 & 13 \\
3 & & 71 & 62 & 22 & 44 & 27 & 21 \\ \hline
\end{tabular}
\pagebreak

\begin{center}
Table Captions
\end{center}

\noindent {\bf Table I}
Energetic and geometric values for several toroidal molecules, according to
H\"uckel's model and to Tersoff's potential. All the molecules have 5 unit
cells. The number of generations $m$, the girths of the first and
last generations $g_1$ and $g_m$ and the length of a ``zig" $z$ determine
the tiling. (See paragraph above Eq.~(4a).)
 The second and third columns are inflations
of the first. $N$ is the number of atoms in the molecule,
 $\epsilon_{h,\ell}$ is the energy of the highest
occupied (resp. lowest unoccupied) molecular orbital, $\bar{\epsilon}$ is the
average energy of delocalized electrons, $\bar{E}$ is the average geometric
energy per atom, $E_{\rm worst}$ is the geometric energy
associated with the least bound atom, and
$\bar{\epsilon}_G=1.575 \beta$ and
$E_G=-7.40{\rm eV}$ are the delocalisation and
the binding energy per atom of an ideal graphitic plane.
The molecule in the last column (denoted by $m=5^1$) is
obtained by a chiral shift $j=1$ of $(m,g_1,g_m,z,n)=(5,2,2,1,5)$

\noindent {\bf Table II} $10^2 \times
(\epsilon_{\ell}-\epsilon_h)/|\beta|$ for molecules with 2 generations
and 5 unit cells, for various values of $g_1, \ g_2$ and $z$.
\newpage

\centerline{Figure Captions}

\noindent {\bf Fig. 1} The torus of revolution and its parametrization. The
dashed curve $C_1$ is a circle of radius $R$ in the $xy$ plane, centered
at the $z$ axis. Through every point $P$ on the surface of the torus
there is a circle $C_2$ of radius $R/\eta$ in a plane that contains the
$z$ axis; the center of $C_2$ belongs to $C_1$. $C_2$ is a longitudinal line.

\noindent {\bf Fig.~2} Solid lines and dashed lines describe two graphs,
which are duals of each other. Both graphs are in the rectangle $0 \le
\theta,\phi \le 2\pi$. We shall draw $\theta$ (resp. $\phi$) as the
vertical (resp. horizontal) coordinate.
 The graph in solid lines has 2 pentagons, 7
hexagons and 2 heptagons. (Periodicity has to be born in mind.) By
construction, at every face of a graph that has $j$ edges, its dual has a
vertex with valence $j$ (where $j$ edges converge).

\noindent {\bf Fig. 3} (a) The two tiles of an odd generation. (b)
The four tiles of an even generation.

\noindent {\bf Fig. 4} Even generation tiling with unnecessary crumpling.

\noindent {\bf Fig. 5} The tiling $m=4, \ g_1=2, \ g_4=z=1$.
$m$ is the number of generations, $g_i$ is the girth of the $i$-th
generation and $z$ is the number of edges in a zig along the (1,2) interface.
 Each tile
has been marked with the number of its generation. The ``X" and the
asterisk are centers
of inversion. The circle at the center of a third generation tile has
been drawn for reference in the text. For clarity, we have
taken a contour that does not cut the tiles (rather than a rectangle);
still, the left and the right boundary are identified, i.~e., this
pattern can be used to cover the plane periodically.

\noindent {\bf Fig. 6} The tiling $m=g_1=g_2=z=2$.

\noindent {\bf Fig. 7} The tiling $m=g_1=g_2=2, \ z=3$. It has the same
number of generations and girths as the tiling in Fig.~6, but longer
``zig" (and, therefore, longer latitudinal lines).

\noindent {\bf Fig. 8} Interfaces between regions of tiles with different
(necessarily consecutive) generation numbers.
 (a) Even generation is higher than the odd generation.
At the interface, the hypotenuse of an even
generation tile matches the cathetus of an odd generation tile.
The interface consists, therefore, of oblique lines.
 (b) Opposite case (odd generation higher than the even generation).
The interface is a horizontal line.

\noindent {\bf Fig.~9} Skeleton of tiling that obeys the 5-6-7 rule and has
more than one stripe. (In the figure, $m=6$ and there are 3 stripes.)
The numbers denote the generation of the tiles in the
region. $g_1$, $g_m$ and $z$ determine completely the unit cell of the
tiling.
The heptagonal vertices of the tiling (not drawn) are at the vertices of
the skeleton where generations 1 and 2 (but not 3) meet. Likewise, the
pentagonal vertices of the tiling are at the vertices of
the skeleton where generations 6 and 5 (but not 4) meet.

\noindent {\bf Fig. 10} The tiling $g_1=2, \ g_2=z=1$, after elongation by 1.

\noindent {\bf Fig. 11} Three unit cells of the tiling $g_1=g_2=z_1=2,
  \ z_2=1$.

\noindent {\bf Fig. 12} Unit cell of the tiling $g_1=2, \ g_2=z=1$, after
a Goldberg inclusion with indices $\ell_1=1, \ \ell_2=2$. The circles
indicate the original vertices before the inclusion.

\noindent {\bf Fig. 13} Tiling of a unit cell of a torus of
Ref.~\cite{Dunlap}. In their notation, this tiling consists of the
tubules $L=M=4$ and $L=8$, $M=0$.

\noindent {\bf Fig. 14} Three-dimensional view of the C$_{480}$ molecule
$(m,g_1,g_2,z,n)=(2,4,4,4,5)$, which minimizes Tersoff's potential. For
clarity, only atoms in the foreground are shown. (a) View from ``above".
(b) Side view.

\noindent {\bf Fig. 15} Same as Fig. 14, for
the C$_{240}$ molecule $(m,g_1,g_5,z,n)=(5,2,2,1,5)$.

\noindent {\bf Fig. 16} Same as Fig. 15, but for a chiral isomer of
C$_{240}$ with chiral shift $j=1$.

\noindent {\bf Fig. 17} Gap in the H\"uckel spectrum for molecules
$(m,g_1,g_2,z,n)=(2,2\ell,2\ell,\ell,5)$, $(2,2\ell,2\ell,\ell,6)$,
$(2,2\ell,2\ell,2\ell,5)$, $(2,2\ell,2\ell,2\ell,6)$,
$(2,2\ell,2\ell,3\ell,5)$, $(2,2\ell,2\ell,3\ell,6)$, for $\ell=1,2,3...$
 For $z/\ell=1$
(resp. 2, 3), calculations were carried up to $\ell=9$ (resp. 8, 5).
The numbers in the graph stand for $z/\ell$. Calculations for $n=6$ were
carried up to $\ell=5$; for the values of $\ell$ for which a number appears
only once, the values of the gap for $n=5$ and $n=6$ are either identical
or indistinguishable within the resolution of the graph. For every $z/\ell$,
$\ell(\epsilon_{\ell}-\epsilon_h)$ exhibits approximate periodicity as a
function of $\ell$, with period 3. Note that the number of atoms in the
molecule is proportional to $\ell^2$

\noindent {\bf Fig. 18} Delocalisation energy for the same molecules studied
in Fig. 17. $\bar{\epsilon}_G=1.575 \beta$
is the delocalisation energy per atom of an ideal graphitic plane.
Again, the ordinates are approximately periodic functions of the
inflation factor $\ell$.

\end{document}